\documentclass[aps,prd,twocolumn,showpacs,floatfix,preprintnumbers]{revtex4}
\usepackage{epsfig}
\def\d{{\rm d}}
\def\sigv{\langle\sigma_{\rm ann}v\rangle}
\newcommand{\lsim}   {\mathrel{\mathop{\kern 0pt \rlap
  {\raise.2ex\hbox{$<$}}}
  \lower.9ex\hbox{\kern-.190em $\sim$}}}
\newcommand{\gsim}   {\mathrel{\mathop{\kern 0pt \rlap
  {\raise.2ex\hbox{$>$}}}
  \lower.9ex\hbox{\kern-.190em $\sim$}}}
\newcommand{\be}{\begin{equation}}
\newcommand{\ee}{\end{equation}}
\newcommand{\bea}{\begin{eqnarray}}
\newcommand{\eea}{\end{eqnarray}}

\begin{document}

\title{Model-independent dark matter annihilation bound from the diffuse
gamma ray flux}
\author{M.~Kachelrie{\ss}$^1$ and P.~D.~Serpico$^2$}
\affiliation{$^{1}$Institutt for fysikk, NTNU, N--7491 Trondheim, Norway\\
$^{2}$Center for Particle Astrophysics, Fermi National Accelerator
Laboratory, Batavia, IL 60510-0500, USA}
\begin{abstract}
An upper limit on the total annihilation cross section of dark
matter (DM) has recently been derived from the observed atmospheric
neutrino background. We show that comparable bounds are obtained for
DM masses around the TeV scale by observations of the diffuse
gamma-ray flux by EGRET, because electroweak bremsstrahlung leads to
non-negligible electromagnetic branching ratios, even if DM
particles only couple to neutrinos at tree level. A better mapping
and the partial resolution of the diffuse gamma-ray background into
astrophysical sources by the GLAST satellite will improve this bound
in the near future.
\end{abstract}
\preprint{FERMILAB-PUB-07-327-A}
\maketitle

%%%%%%%%%%%%%%%%%%%%%%%%%%%%%%%%%%%%%%%%%%%%%%%%%%%%%%%%%%%%
\section{Introduction}
%%%%%%%%%%%%%%%%%%%%%%%%%%%%%%%%%%%%%%%%%%%%%%%%%%%%%%%%%%%%
One promising way to detect dark matter (DM) is indirectly via its
annihilation (or decay) products. The DM annihilation
products—barring exotic models with additional stable and relatively
light particles—are standard model (SM) particles, although with
model-dependent branching ratios. Using atmospheric neutrino data,
the authors of Ref.~\cite{Beacom:2006tt} derived an observational
upper bound on the annihilation cross section $\sigv$ of any DM
candidate, assuming that it annihilates only into the least
detectable final states in the SM, namely neutrinos. Allowing only
couplings to neutrinos might be not only a conservative assumption
needed to derive this bound, but could be realized in nature:
Possible DM candidates, like the Majoron, with this property exist.
Moreover, the bound on the diffuse gamma-ray background from EGRET
observations~\cite{Hunter:1997,Sreekumar:1997un,Strong:2004ry}
translates into extremely restrictive limits on the branching ratios
in electromagnetic and hadronic DM annihilation channels. Therefore,
models with high annihilation rates proposed to solve the ''cusp
problem'' of conventional cold DM (see
e.g.~\cite{Kaplinghat:2000vt}) and DM masses $m_X$ above ${\cal
O}({\rm GeV})$ are likely to require either fine-tuning or should
couple the DM particle only to neutrinos.

The latter possibility has already been invoked in exotic scenarios
explaining the origin of ultrahigh energy cosmic rays.
Reference~\cite{Gelmini:1999ds}proposed that supermassive relic
particles decay only into neutrinos, thereby contributing to the
ultrahigh energy cosmic ray flux through the $Z$ burst mechanism and
avoiding at the same time constraints from the diffuse gamma-ray
background. However, the authors of Ref.~\cite{Berezinsky:2002hq}
showed that electroweak jet cascading leads to a non-negligible
electromagnetic branching ratio and rules out these models.

In this work, we extend this argument to annihilating dark matter of
lower mass, showing that this mechanism combined with the limit on
the diffuse gamma radiation by the EGRET satellite provides
competitive observational constraints on $\sigv$ for masses around
the TeV scale. Future observations of the diffuse gamma-ray flux by
the GLAST satellite should improve these bounds. We also find that
the strongest and most robust way to constrain $\sigv$ is to use the
DM signal associated with the galactic halo, instead of the diffuse
flux from cosmologically distributed dark matter. We comment on the
possibility to improve the neutrino bounds as well by exploiting the
strongly peaked angular distribution expected from annihilations in
the galactic dark-matter halo. In Sec. \ref{DMinput}, we discuss the
properties of dark matter relevant here, while Sec. \ref{background}
is devoted to the data used to derive the bound. The bound is
presented and commented upon in Sec. \ref{boundSec}. In Sec.
\ref{summary}, we discuss possible improvements and finally
conclude.

%%%%%%%%%%%%%%%%%%%%%%%%%%%%%%%%%%%%%%%%%%%%%%%%%%%%%%%%%%%%%
\section{The dark matter input} \label{DMinput}

In Ref.~\cite{Beacom:2006tt}, the expected dominating contribution
to the diffuse neutrino flux was estimated from the integrated
extragalactic contribution to dark matter annihilations, and
compared with the measured atmospheric neutrino flux. Unfortunately,
the extragalactic flux strongly depends on the shape of dark matter
halos and their degree of clumpiness. A robust estimate is thus
difficult to achieve. Although in Ref.~\cite{Beacom:2006tt} a
relatively modest value of $2\times 10^{5}$ for the enhancement due
to the clumpiness of DM was used, even values lower by a factor of
$\cal{O}$(10) are possible. To be more conservative, we use the
diffuse photon flux from the {\em smooth\/} DM distribution in the
halo of our Galaxy since: {\em (i)\/} Its normalization and
distribution is better known (within a factor $\sim 2$); {\em
(ii)\/} It is truly a lower limit for the DM annihilation flux
\cite{Hooper:2007be}. Substructure in our halo is expected to
augment it up to orders of magnitude (see e.g. the parametric study
\cite{Hooper:2007be} for our Galaxy or the study
\cite{Strigari:2006rd} for dwarf galaxy satellites). Note that the
contribution from the diffuse extragalactic photon background from
DM annihilations further enhances the total DM emission. By
neglecting both the substructure in our halo and the extragalactic
contribution, we are being conservative.

The differential flux of photons from dark matter annihilations is
\footnote{Assuming non self-conjugated particles, an additional
factor 1/2 is needed.}
\begin{equation}
I_{\rm
sm}(E,\psi)=\frac{\d N_\gamma}{\d E}\,\frac{\sigv}{2\,m_X^2}\,\int_{\rm
l.o.s.} \d s\,\frac{\rho_{\rm sm}^2[r(s,\psi)]}{4\pi}, \label{Ism}
\end{equation}
where $r(s,\psi)=(r_\odot^2+s^2-2\,r_\odot\,s\cos\psi)^{1/2}$,
$\psi$ is the angle between the direction in the sky and the
galactic center (GC), $r_\odot\approx 8.0\,$kpc is the solar
distance from the GC, and $s$ the distance from the Sun along the
line-of-sight (l.o.s.). In terms of galactic latitude $b$ and
longitude $l$, one has
\begin{equation}
\cos\psi=\cos b\cos l\,. \label{psi}
\end{equation}
Particle physics enters via the DM mass
$m_X$, the annihilation cross section $\sigv$, and
the photon differential energy spectrum $\d N_\gamma/\d E$
per annihilation.
Concerning the DM halo profile, we adopt for the smooth DM mass density
$\rho_{\rm sm}$ a Navarro-Frenk-White
profile~\cite{NFW}
\begin{equation} \rho_{\rm
sm}(r)=\rho_\odot\left(\frac{r_\odot}{r}\right)
\left(\frac{r_\odot+a}{r+a}\right)^2,
\end{equation}
where we choose $\rho_\odot =0.3\,$GeV/cm$^3$ as the dark matter
density at the solar distance from the GC, and $a= 45\,$kpc as the
characteristic scale below which the profile scales as $r^{-1}$. The
galactic halo DM flux has a significant angular dependence, with
possibly large fluxes from the galactic center region. However, the
DM profile in the inner regions of the Galaxy is highly uncertain.
To be conservative, we shall only use the NFW profile for
$r>1\,$kpc, a region where numerical simulations of DM halos have
reached convergence and the results are robust
\cite{Stoehr:2003hf,Diemand:2006ik}. Of course, other choices for
the profile are possible, but all of them agree in the range of
distances considered here, differing primarily in the central region
of the halos. Since here we are focusing on the galactic diffuse
emission rather than that from the GC, the residual uncertainties
which are introduced through the choice of profile (a factor $\sim
2$) are negligible for our discussion.

%%%%%%%%%%%%%%%%%%%%%%%%%%%%%%%%%%%%%%%%%%%%%%%%%%%%%%%%%%%%
\section{The diffuse gamma ray backgrounds}\label{background}
\begin{figure}[!tb]
\hspace*{-2.3cm} \epsfig{file=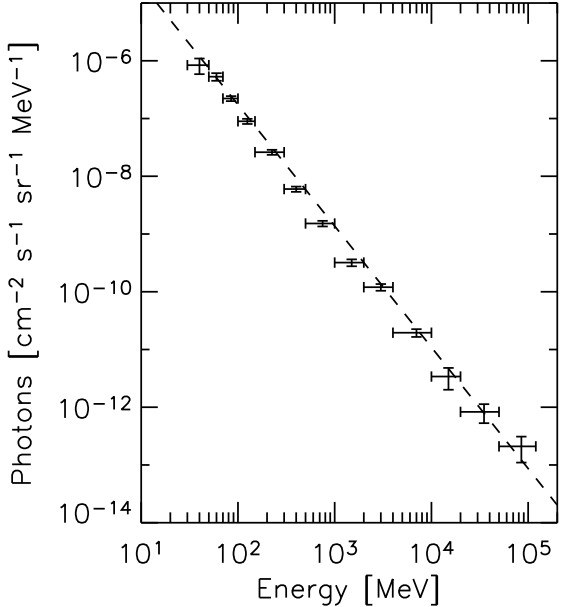,width=1.8\columnwidth}
\caption{EGRET data for the diffuse extragalactic gamma ray flux,
according to~\cite{Strong:2004ry}, and the fit of the original
analysis in~\cite{Sreekumar:1997un}. \label{EGRETspectrum}}
\end{figure}

%%%%%%%%%%%%%%%%%%%%%%%%%%%%%%%%%%%%%%%%%%%%%%%%%%%%%%%%%%%%
The overall diffuse gamma-ray radiation can be qualitatively divided
into a galactic and an extragalactic contribution. Since the latter
is not simply the isotropic part of the flux,  the separation of
these two components can be done at present only assuming a specific
model for the production of secondaries by cosmic rays in the
galactic disk and halo. (However, a measurement of the cosmological
Compton-Getting effect that should be achievable for GLAST would
provide a model-independent way to separate the two
contributions~\cite{Kachelriess:2006aq}). A significant fraction of
the quasi-isotropic component, especially in the GeV range, may be
due to high-latitude galactic emission coming from processes in the
magnetized halo of the Milky Way. For our purposes here, a detailed
analysis is not required, and thus we employ a fit of the galactic
diffuse flux proposed in~\cite{Bergstrom:1997fj} and calibrated on
EGRET data around the GeV~\cite{Hunter:1997},
\begin{widetext}
\be \label{spectrumGal} I_{\rm gal}(E)= N_0(l,b) \times 10^{-6}
\left(\frac{E}{{\rm GeV}}\right) ^{-2.7}\,{\rm cm}^{-2}{\rm
s}^{-1}{\rm sr}^{-1}{\rm GeV}^{-1} \,, \ee
where the arguments are in degrees, $-180^\circ\leq l\leq 180^\circ$
and $-90^\circ\leq b\leq 90^\circ$,
\begin{equation}   \label{linsys1}
N_{0}(l,b)=\left\{
\begin{array}{cc}
\frac{85.5}{\sqrt{1+(l/35)^2}\sqrt{1+[b/(1.1+0.022\,|l|)]^2}}+0.5 \,,
&\:\:|l|\geq 30^{\circ}\\
\frac{85.5}{\sqrt{1+(l/35)^2}\sqrt{1+(b/1.8)^2}}+0.5 \,,
&\:\:|l|\leq 30^{\circ}
\end{array}
\right..
\end{equation}

The EGRET collaboration derived the intensity of the extragalactic
gamma-ray flux as~\cite{Sreekumar:1997un}
\begin{equation}
I_{\rm ex}(E)=(7.32\pm 0.34)\times 10^{-6}\left(\frac{E}{0.451 {\rm
GeV}}\right) ^{-2.10\pm0.03} {\rm cm}^{-2}{\rm s}^{-1}{\rm
sr}^{-1}{\rm GeV}^{-1} \, , \label{spectrum98}
\end{equation}
valid from $E\sim\,$10~MeV to $E\sim\,$100~GeV.
\end{widetext}
The reanalysis of the data
performed in~\cite{Strong:2004ry}, based on a revised model for the
galactic propagation of cosmic rays, deduced an extragalactic
spectrum significantly lowered with respect to
Eq.~(\ref{spectrum98}) at intermediate energies, while closer to the
original result of Eq.~(\ref{spectrum98}) at the lowest and highest
energy points. In Fig.~\ref{EGRETspectrum}, we show the points
according to this reevaluation, together with the fit of
Eq.~(\ref{spectrum98}).
To derive our constraint, we shall ask that the photon flux from DM
annihilations, integrated in each of the energy bins of
Fig.~\ref{EGRETspectrum} and in the whole energy range covered by
EGRET, remains below the sum of the upper limit for the
extragalactic flux plus the galactic emission estimated according to
the fit of Eq. (\ref{spectrumGal}). Note that since we compare the
signal with the sum of the two contributions, the precise
extragalactic fraction of the diffuse radiation is basically
irrelevant. To be conservative, we shall compare the DM photon flux
to the background profiles along the curve $l=0$, since the galactic
background is maximum at this longitude (see
Eq.~(\ref{spectrumGal})).

%%%%%%%%%%%%%%%%%%%%%%%%%%%%%%%%%%%%%%%%%%%%%%%%%%%%%%%%%%%%%
\section{Gamma-ray emission from DM annihilation into neutrinos}
\label{boundSec}
%%%%%%%%%%%%%%%%%%%%%%%%%%%%%%%%%%%%%%%%%%%%%%%%%%%%%%%%%%%%%
%
\begin{figure}[!thb]
\hspace*{-0.6cm}
\epsfig{file=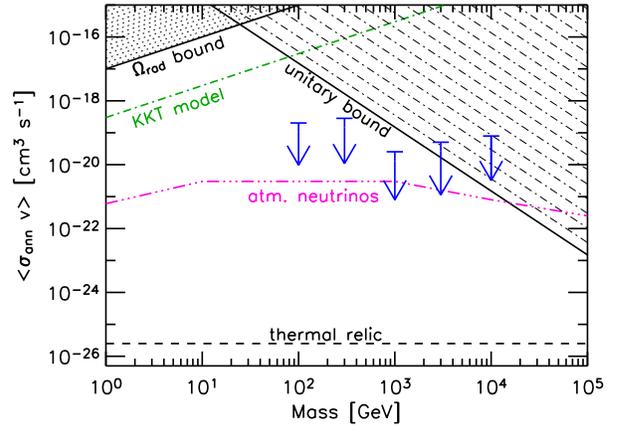,width=1.25\columnwidth}
\caption{Bounds on $\sigv$ versus $m_X$ from diffuse $\gamma$ rays
  (blue arrows), atmospheric neutrino data~\cite{Beacom:2006tt}
  (magenta line) together with the expectation for a thermal relic
  (for s-wave annihilation), the KKT model and the unitary limit. See
  the text for details.
\label{bound}}
\end{figure}
By assumption, the DM particles $X$ couple on tree-level only to
neutrinos. Hence the only possible $2\to2$ annihilation process is
$X X\to \bar\nu\nu$ with an unspecified intermediate state that has
negligible couplings to SM particles. Then the dominant $2\to 3$ and
$2\to 4$ processes are the bremsstrahlung of an electroweak gauge
boson that subsequently decays: $X X\to \bar\nu\nu Z, \nu e^\pm
W^\mp$ and $X X\to \bar\nu\nu\bar ff$. The branching ratio
$R=\sigma(X X\to \bar\nu\nu Z)/\sigma(X X\to \bar\nu\nu)$ depends
generally only for $Q^2\sim m_X^2$ on the details of the underlying
$2\to2$ process. One can distinguish three different regimes of this
process: $i)$ the Fermi regime $m_X\lsim m_Z$ with ${\cal O}(R)=
[\alpha_2/(4\pi)]^2(m_X/m_Z)^4$, $ii)$ the perturbative electroweak
regime $m_Z~\lsim~m_X~\lsim~\alpha_2/(4\pi)\ln^2(m_X/m_Z)^2\sim
10^6\,$GeV where $R$ grows from ${\cal O}(\alpha_2/(4\pi))$ to
${\cal O}(0.1)$, and $iii)$ the non-perturbative regime where large
logarithms over-compensate the small electroweak coupling
$\alpha_2$~\cite{Berezinsky:2002hq}. Here, we consider regime $ii)$
and can use therefore standard perturbation theory for the
evaluation of $R$. Numerical values of $R$ are given in Tab.~1.

\begin{table}[b]
\caption{The branching ratio
$R=\sigma(X X\to \bar\nu\nu Z)/\sigma(X X\to \bar\nu\nu)$ as function
of $m_X$.}
\begin{tabular}{|c||c|c|c|c|c|}
\hline
$m_X$/GeV       & $100$ & $300$ & $1000$ & $3000$&  $10^{4}$
 \\\hline
$R$/\% & 0.01    &  0.02    &  0.87      & 1.9       & 3.4\\
\hline
\end{tabular}
%\vskip0.3cm
\end{table}

The dominant source of photons are $\pi^0$ produced in quark jets from
$W$ and $Z$ decays. The resulting differential photon energy spectrum
$\d N_\gamma/\d E$ has been simulated using HERWIG~\cite{H}.

The obtained bound from the EGRET limit is shown in Fig.~\ref{bound}
with arrows together with the limit from Ref.~\cite{Beacom:2006tt}
using atmospheric neutrino data. The upper extreme of the arrow
indicates the bound obtained by comparing the emissions at the
highest galactic latitudes ($b=\pi/2$, $l=0$), while the lower
extreme is the bound coming from the inner Galaxy emission ($b=1/8$,
$l=0$). The length of the arrow thus quantifies the improvement due
to our simple, angular-dependent analysis. Indicated are also the
required value for a standard thermal relic with an annihilation
cross section dominated by the s-wave contribution, $\sigv\approx
2.5\times 10^{-26}$cm$^3$/s, the unitary limit $\sigv\leq
4\pi/(v\,m_X^2)$ for $v=300\,$km/s, appropriate for the Milky way,
and the constraints on the cosmological relativistic energy density
from \cite{Zentner:2001zr}.

%%%%%%%%%%%%%%%%%%%%%%%%%%%%%%%%%%%%%%%%%%%%%%%%%%%%%%%%%%%%%
\section{Discussion and conclusion}\label{summary}
%%%%%%%%%%%%%%%%%%%%%%%%%%%%%%%%%%%%%%%%%%%%%%%%%%%%%%%%%%%%
In this paper we have shown that, even if dark-matter particles
annihilate at tree-level only into neutrinos, diffuse gamma-ray data
provide interesting constraints on their annihilation cross section
because of electroweak bremsstrahlung. These bounds are comparable
to the atmospheric neutrino bound from Ref.~\cite{Beacom:2006tt} in
the mass range between $\sim 100\,$GeV and the onset of the stronger
unitary bound around 10 TeV. Any appreciable branching ratio at tree
level in electromagnetically interacting particles would lead to
much stronger constraints from gamma-rays, but they are not as
conservative as the bounds derived here or in
Ref.~\cite{Beacom:2006tt}. A major improvement in the gamma-ray
bound is expected from the GLAST satellite \cite{glast}, to be
launched by the beginning of 2008. In particular, GLAST should
resolve most of the diffuse flux of astrophysical origin, and map
both the galactic and extragalactic diffuse emission with much
higher accuracy, thereby improving  the bound derived here. On the
other hand, our results also suggest that the neutrino bound may be
tightened as well by considering the DM annihilation in the galactic
halo and taking into account the strong angular dependence on the
halo signal \footnote{ After this paper had been submitted, an
improved neutrino bound has been presented in \cite{Yuksel:2007ac},
obtained actually from an angular-dependent treatment of the halo
signal, confirming our expectations.}.

As a further application of our results, we note that the electroweak
higher-order corrections discussed here also contribute to increase
the robustness of the bounds on strongly interacting dark matter from
the Earth's heat flow in Ref.~\cite{Mack:2007xj}. Above the TeV scale,
electroweak bremsstrahlung put a lower bound of $\mathcal{O}(1\%)$ on
the energy released in other-than-neutrino channels, thus guaranteeing
that an appreciable energy is release by annihilations in the interior
of the Earth even for models with tree-level annihilations in
neutrinos only.

\section*{Acknowledgements}
We are grateful to John Beacom, Mike Seymour and Brian Webber for
helpful comments. PS acknowledges support by the US Department of
Energy and by NASA grant NAG5-10842.

\end{document}